\documentstyle[aps,epsfig,prl]{revtex}

\begin{document}
\title{{\LARGE Magnetization of Nuclear-Spin-Polarization-Induced Quantum Ring}}
\author{{S. N. Shevchenko}$^{1,2}${\it , }\-{\it Yu. V. Pershin}$^{1,3}${\it , and
I. D. Vagner}$^{1,3,4}$}
\address{$^{1}$Grenoble High Magnetic Fields Laboratory, Max-Planck-Institute f\"{u}r
Festkorperforschung and CNRS, BP 166, F-38042 Grenoble Cedex 9,
France\\ $^{2}$B. I. Verkin Institute for Low Temperature Physics
and Engineering,\ 47 Lenin Avenue, 61103 Kharkov, Ukraine\\
$^{3}$Center for Quantum Device Technology, Department of Physics
and Department of Electrical and Computer Engineering, Clarkson
University, Potsdam, NY 13699-5820, USA\\ $^{4}$Holon Academic
Institute of Technology, 52 Golomb St., Holon 58102, Israel}
\maketitle

\bigskip

\begin{abstract}
Properties of a Nuclear-Spin-Polarization-Induced Quantum Ring
(NSPI QR) are studied theoretically. In the proposed system a
local nuclear spin polarization creates an effective hyperfine
field which confines the electrons with the spins opposite to the
hyperfine field to the regions of maximal nuclear spin
polarization. We investigate the electron energy spectrum and the
magnetic response of NSPI QR and their evolution in time due to
the nuclear spin diffusion and relaxation.
\bigskip

\end{abstract}


Nuclear spin polarization in semiconductor heterostructures has
several intriguing properties such as the large value of effective
nuclear hyperfine field, experienced by electrons, and long
longitudinal nuclear spin relaxation time
\cite{Paget,Salis,Dixon,Berg}. Nuclear spins embedded into
mesoscopic systems were proposed to be used in spintronic devices
and for quantum computation (see \cite{devices,PVW2003} and
references therein). There are several techniques of polarizing
nuclear spins. The most extensively employed methods use the
optical pumping scheme \cite{Tycko} or transport polarization
scheme \cite{Kane and Wald}. These techniques are developed to
allow local polarization and controllability of nuclear spins
\cite{recent}.

It has been recently suggested that locally polarized nuclei can
be used to confine the electrons of two-dimensional electron gas
(2DEG) into different so-called nuclear-spin-polarization-induced
structures \cite{our paper}-\cite{LMKHV03}. In these papers the
electronic states in NSPI quantum wire \cite{our paper}, quantum
dot \cite{FIPV01}-\cite{Yura}, and periodic structure
\cite{LMKHV03} were studied. The idea of producing NSPI structures
was as following. The local nuclear spin polarization can be
described by the effective hyperfine field ${\bf B}_{hf}$, which
acts at electronic spins and enters in electronic
Hamiltonian through Zeeman-type potential $\frac{g}{2}\mu _{B}{\bf \sigma B}%
_{hf}\left( {{\bf r},t}\right) $ \cite{Slichter}. (Here $g$ is the electron $%
g$-factor, $\mu _{B}$ is the Bohr magneton and ${\bf \sigma }$ is the
Pauli-matrix vector corresponding to the electron spin.) The Zeeman
splitting results in that the potential is attractive for electrons with one
spin projection and repulsive for others. The energy is proposed to be
shifted by the constant potential of the order of the 2DEG Fermi energy ($%
eV_{gate}\simeq E_{F}^{2DEG}$) by the means of the gate located under the
2DEG. Then the potential landscape, created by the inhomogeneous nuclear
spin polarization, can be considered as the confining potential

\begin{equation}
U_{conf}({\bf r}{,t})=-\frac{\left| g\right| }{2}\mu _{B}B_{hf}\left( {{\bf r%
},t}\right) .  \label{U_conf}
\end{equation}
It is assumed that $B_{hf}\left( {{\bf r},t}\right) >0$ in Eq. (\ref{U_conf}%
). The maximum nuclear field in GaAs can be as high as
$B_{hf}=$5.3T in the limit that all nuclear spins are fully
polarized \cite{Paget}. This high level of nuclear spin
polarization has been achieved experimentally. For example, the
optical pumping of nuclear spins in 2DEG has demonstrated nuclear
spin polarization of the order of $90\%$, \cite{Salis}. A
similarly high polarization has been created by quantum Hall edge
states ($85\%$) \cite{Dixon}.

In this paper we extend the study of NSPI systems by considering the
electronic structure of the NSPI QR. This means that we assume the hyperfine
field in Eq.(\ref{U_conf}) to be axially symmetric and to have maximum at $%
r\simeq r_{0}$ (which can be considered as the characteristic radius of the
ring). The energy profile of the system is schematically presented in Fig.
1. The system is placed into an external magnetic field ${\bf B}$
perpendicular to the ring plane, i.e. along $z$-axis. It is well known that
in such geometry an equilibrium persistent current appears as a direct
manifestation of Aharonov-Bohm effect (for a review see \cite{ABE and PC},
\cite{TI}). The persistent current yields the orbital magnetic moment, which
is an oscillating function of the Aharonov-Bohm flux. In the present paper
we calculate the magnetic moment of NSPI QR, which may be directly measured
in experiments.

We would like to underline the difference of the present
investigation from previous works in this field. In Ref.
\cite{VRWZ98} the interplay of nuclear magnetization and
persistent current was considered. The authors of \cite{VRWZ98}
have shown that the Aharonov-Bohm-like oscillations of the
persistent current in mesoscopic rings could exist at zero
magnetic field as the result of combined action of the
nonequilibrium nuclear spin population and charge carrier
spin-orbit interaction. The role of the nuclear spin magnetization
in Ref. \cite{VRWZ98} and in the present paper is different. In
NSPI QR the local nuclear spin polarization is used only to
confine electrons into a ring and the oscillations of the magnetic
moment are because of the evolution of the confining potential,
while in Ref. \cite{VRWZ98} a time-dependent non-trivial profile
of nuclear spin polarization enters into the phase of the electron
wave function and causes Aharonov-Bohm-like oscillations.

Local nuclear spin polarization is not static. Processes of
nuclear spin diffusion and relaxation change the hyperfine field
in time making the electron confining potential
(Eq.(\ref{U_conf})) time-dependent. To find the hyperfine magnetic
field $B_{hf}\left( r,t\right) $ we should solve the diffusion
equation with relaxation. We do this as in Refs. \cite{our paper}
and \cite{Yura} assuming that the nuclear hyperfine field has
Gaussian distribution at the initial moment of time $t=0$. The
solution of the diffusion equation with relaxation has a form

\begin{equation}
B_{hf}\left( r,t\right) =B_{hf}\left( r_{0},0\right) e^{-\frac{t}{T_{1}}%
}\left( 1+\frac{t}{t_{0}}\right) ^{-1}e^{-\frac{r^{2}+r_{0}^{2}}{%
2d_{0}^{2}\left( 1+\frac{t}{t_{0}}\right) }}I_{0}\left( \frac{rr_{0}}{%
d_{0}^{2}\left( 1+\frac{t}{t_{0}}\right) }\right) \text{ \ \ ,}
\label{B(r,t)}
\end{equation}
where $t_{0}=\frac{d^{2}}{2D}$ is a characteristic diffusion time; $D$ is
the nuclear spin diffusion coefficient; $T_{1}$ is the nuclear spin
relaxation time; $r_{0}$, $d_{0},$ and $B_{hf}\left( r_{0},0\right) $ define
the radius, the half-width and the amplitude of the distribution of the
hyperfine field at $t=0$, respectively; $I_{0}(z)$ is the modified Bessel
function. When the width of the ring is smaller than its radius ($2d_{0}%
\sqrt{1+t/t_{0}}<r_{0}$), the approximate expression for the nuclear
hyperfine field is:

\begin{equation}
B_{hf}\left( r,t\right) \simeq B_{hf}\left( r_{0},0\right) \frac{\exp
(-t/T_{1})}{\sqrt{1+t/t_{0}}}\sqrt{\frac{r_{0}}{r}}\exp \left( -\frac{%
(r-r_{0})^{2}}{2d_{0}^{2}(1+t/t_{0})}\right) .  \label{B_hypf-Gauss}
\end{equation}
This time-dependent hyperfine magnetic field corresponds (due to Eq.(\ref
{U_conf})) to the confining potential with minimum at $r\simeq r_{0}$ , the
half-width
\[
d(t)\simeq d_{0}\sqrt{1+t/t_{0}}
\]
and the depth
\[
V_{0}(t)=-\frac{\left| g\right| }{2}\mu _{B}B_{hf}\left( r_{0},t\right)
\simeq -\frac{\left| g\right| }{2}\mu _{B}B_{hf}\left( r_{0},0\right) \frac{%
\exp (-t/T_{1})}{\sqrt{1+t/t_{0}}}.
\]

Since the characteristic time of nuclear spin evolution is much
larger than the time scale of electron motion in the mesoscopic
ring, we can consider the electrons in the quasi-time-independent
confining potential $U_{conf}\left( r,t \right)$ at each moment of
time. To proceed, we approximate the exact hyperfine field
potential (Eq.(\ref{U_conf})) by the following potential

\begin{equation}
\widetilde{U}_{conf}(r)=a_{1}r^{-2}+a_{2}r^{2}+a_{3}  \label{Uconfmod}
\end{equation}
where $a_{i}$ are time-dependent parameters as described below. Such a
potential was a success to model the electronic states in both quantum dots
and rings\cite{TI}. To make this model potential (Eq.(\ref{Uconfmod}))
corresponding to the confining potential of NSPI QR, given by Eqs. (\ref
{U_conf}) and (\ref{B_hypf-Gauss}), we adjust the position of the minimum,
depth and width of these potentials. The following dependence of the
coefficients $a_{i}$ on the parameters of the latter potential is obtained: $%
a_{1}=-V_{0}r_{0}^{4}/8d^{2}$, $a_{2}=-V_{0}/8d^{2}$, and $a_{3}=V_{0}\left(
1+r_{0}^{2}/4d^{2}\right) $.

The solution of the Schr\"odinger equation with approximating potential (\ref
{Uconfmod}) can be expressed in terms of the hypergeometric function \cite
{LL}, \cite{TI} and the energy spectrum is

\begin{equation}
E_{n,m}=\left( n+\frac{1}{2}+\frac{M}{2}\right) \hbar \omega +\frac{m}{2}%
\hbar \omega _{c}+V_{0}\left( 1+\frac{r_{0}^{2}}{4d^{2}}\right) ,
\label{spectrum}
\end{equation}
where we used the following notations: $M=\sqrt{m^{2}-\frac{V_{0}}{E_{0}}%
\frac{r_{0}^{2}}{8d^{2}}}$, $\omega =\sqrt{\omega _{c}^{2}+\omega _{0}^{2}}$%
, $E_{0}=\frac{\hbar ^{2}}{2m^{\ast }r_{0}^{2}}$, $\omega _{c}=\left|
e\right| B/m^{\ast }c$ is the cyclotron frequency, $m^{\ast }$ and $e$ are
the electron effective mass and charge, $n=0,1,2,...$ and $m=0,\pm 1,\pm
2,...$, $\omega _{0}=\left( -\frac{V_{0}}{m^{\ast }d^{2}}\right) ^{1/2}$
stands for the frequency of the parabolic potential to which $\widetilde{U}%
_{conf}$ tends at $r\sim r_{0}$ ($\left. \frac{d^{2}\widetilde{U}_{conf}}{%
dr^{2}}\right| _{r=r_{0}}=m^{\ast }\omega _{0}^{2}$). In the limit of
one-dimensional ring ($\hbar \omega _{0}\gg \hbar \omega _{c},$ $E_{0}$) we
obtain from Eq.(\ref{spectrum}) for $n=0$: $E_{n,m}\simeq V_{0}+\frac{\hbar
\omega _{0}}{2}+E_{0}\left( m+\frac{\Phi }{\Phi _{0}}\right) ^{2}$, where $%
\Phi $\ is the flux of external magnetic field, $\Phi _{0}=hc/e$. The
spectrum of NSPI quantum dot can be obtained from Eq.(\ref{spectrum}) in the
limit when the inner ring radius tends to zero. For example in the weak
magnetic field, i.e. assuming $E_{0}\gg \hbar \omega _{0}\gg \hbar \omega
_{c}$, we get: $E_{n,m}\simeq 2V_{0}+\hbar \omega _{0}(n+\frac{1}{2})+\frac{%
\hbar \omega _{0}}{2}\left( \left| m\right| +\frac{\omega _{c}}{\omega _{0}}%
m\right) $.

Further we study the magnetic moment of the electronic system which is given
by the standard formula:
\begin{equation}
{\cal M}=\sum_{n,m}f_{F}(E_{n,m}){\cal M}_{n,m},
\end{equation}
where $f_{F}$ is the Fermi distribution function and the magnetic moment of $%
(n,m)-$ state is
\begin{equation}
{\cal M}_{n,m}=-\frac{\partial E_{n,m}}{\partial B}=M_0\left(
m+\left( M+2n+1\right) \frac{\omega _{c}}{\omega }\right) ,
\end{equation}
where $M_0=-\frac{e \hbar}{2m^*}$.

If we are interested in the magnetic field dependence of the
magnetic moment, we would obtain the results thoroughly presented
in Ref.\cite{TI}. Besides, in NSPI QR there is the time evolution
of the nuclear spin polarization due to diffusion and relaxation.
The time evolution of the electronic structure of NSPI QR depends
upon the relation between two characteristic times, $T_{1}$ and
$t_{0}$. Correspondingly three characteristic time-regimes can be
discussed as in Ref. \cite{our paper}. The illustrations of the
obtained results are shown on Fig. 2 for the case of ''relaxation
regime'' when $T_{1}<2t_{0}$. We present the time evolution of
the energy spectrum $E_{n,m}$ (Fig. 2a), number of electrons $%
N=\sum_{n,m}f_{F}(E_{n,m})$ (Fig. 2b),\ and the magnetic moment
${\cal M}$ (Fig. 2c) in NSPI QR. We have considered GaAs-AlGaAs
heterostructure based
NSPI QR assuming ring radius $r_{0}=1\mu m$, initial NSPI QR half-width $%
d_{0}=0.1\mu m$ and initial value of hyperfine field $
B_{hf}(r_{0},0)=2.65T$ (50\% of nuclear spins are polarized). The
corresponding energies are $E_{0}=0.57\cdot 10^{-3}meV$ and $%
V_{0}(0)=-3.4\cdot 10^{-2}meV$.

Let us consider the results of our calculations in more detail. In
the relaxation regime presented here, the change of the confining
potential is described by the first exponential factor in Eq.
(\ref{B_hypf-Gauss}). The amplitude of the confining potential
exponentially decreases in time, while the shape only slightly
changes due to the diffusion. Consequently, the depth of the
energy levels and their number decreases with time (Fig. 2a,b). As
the result, the magnetization of electrons in NSPI QR changes its
value and sign with time (Fig. 2c). Each sign change of the
magnetization occurs when the number of the electron in systems
decreases by one. The dependence of the energy spectrum $E_{n,m}$,
number of electrons $N$,\ and the magnetic moment ${\cal M}$ on
time is characterized by the parameters, which define the value
and the profile of the initial nuclear polarization, by the value
of external magnetic flux, and by the parameters of the electronic
subsystem.

In conclusion, we have studied the electronic spectrum and
magnetization of NSPI QR. In this structure the spin-polarized
electrons are confined by locally polarized nuclear spins. The
most interesting feature of NSPI QR is that its properties are
time-dependent at fixed values of external parameters, such as,
for example, the applied magnetic field. This effect is due to the
nuclear spin diffusion and relaxation processes. Experimental
realization of NSPI QR will demonstrate a new kind of mesoscopic
phenomenon and can be useful in future spintronic devices. On the
other hand, investigation of magnetic moment could be used to
extract information on nuclear spin relaxation and diffusion
mechanisms.

This research was supported in part by the National Science
Foundation, grant DMR-0121146. S. N. Sh. acknowledges the
hospitality of Grenoble High Magnetic Field Laboratory and thanks
Yu. A. Kolesnichenko for helpful discussions.

\newpage

\begin{figure}[tbp]
\centering
\includegraphics[width=8cm, angle=270]{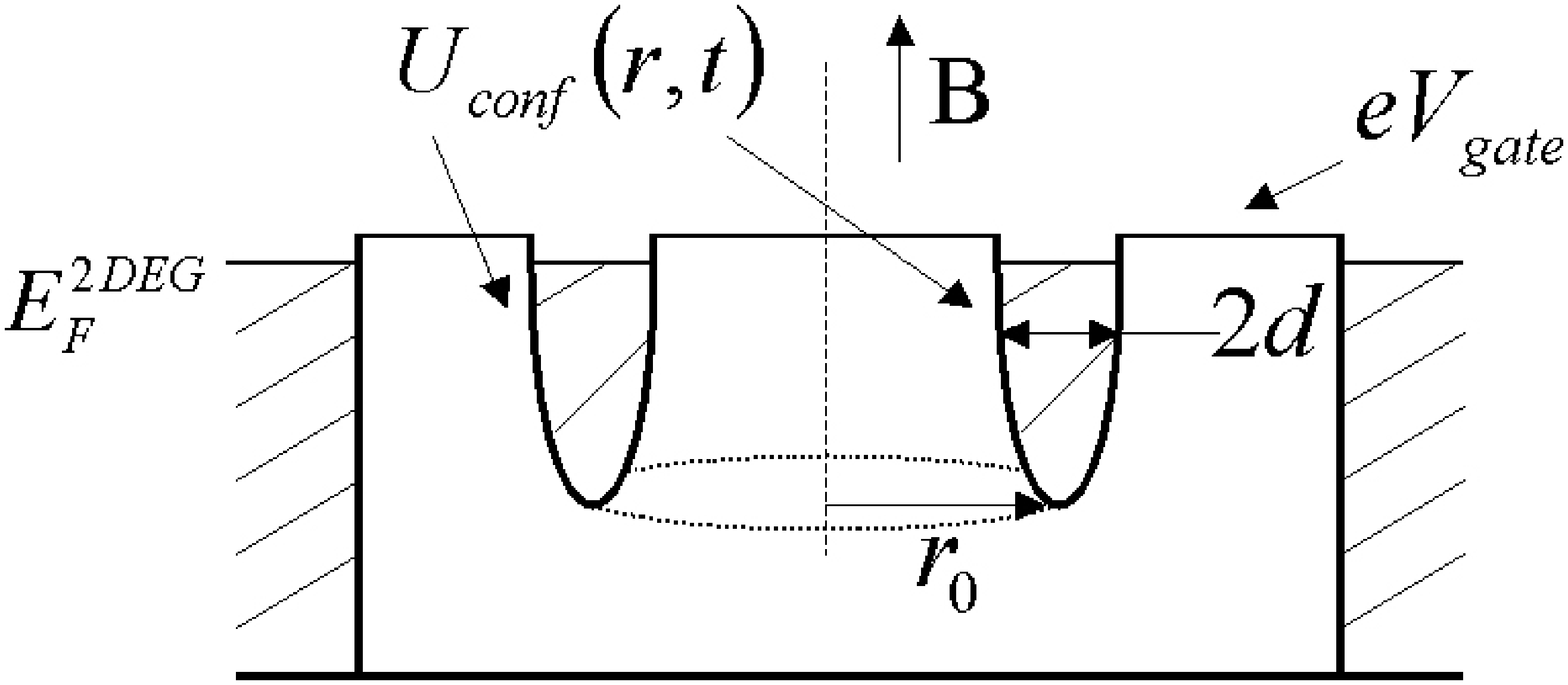}
\bigskip
\caption{Schematic representation of NSPI QR energy structure. The
shaded regions are occupied by electrons.} \label{en_diff}
\end{figure}

\newpage

\begin{figure}[tbp]
\centering
\includegraphics[width=10cm, angle=0]{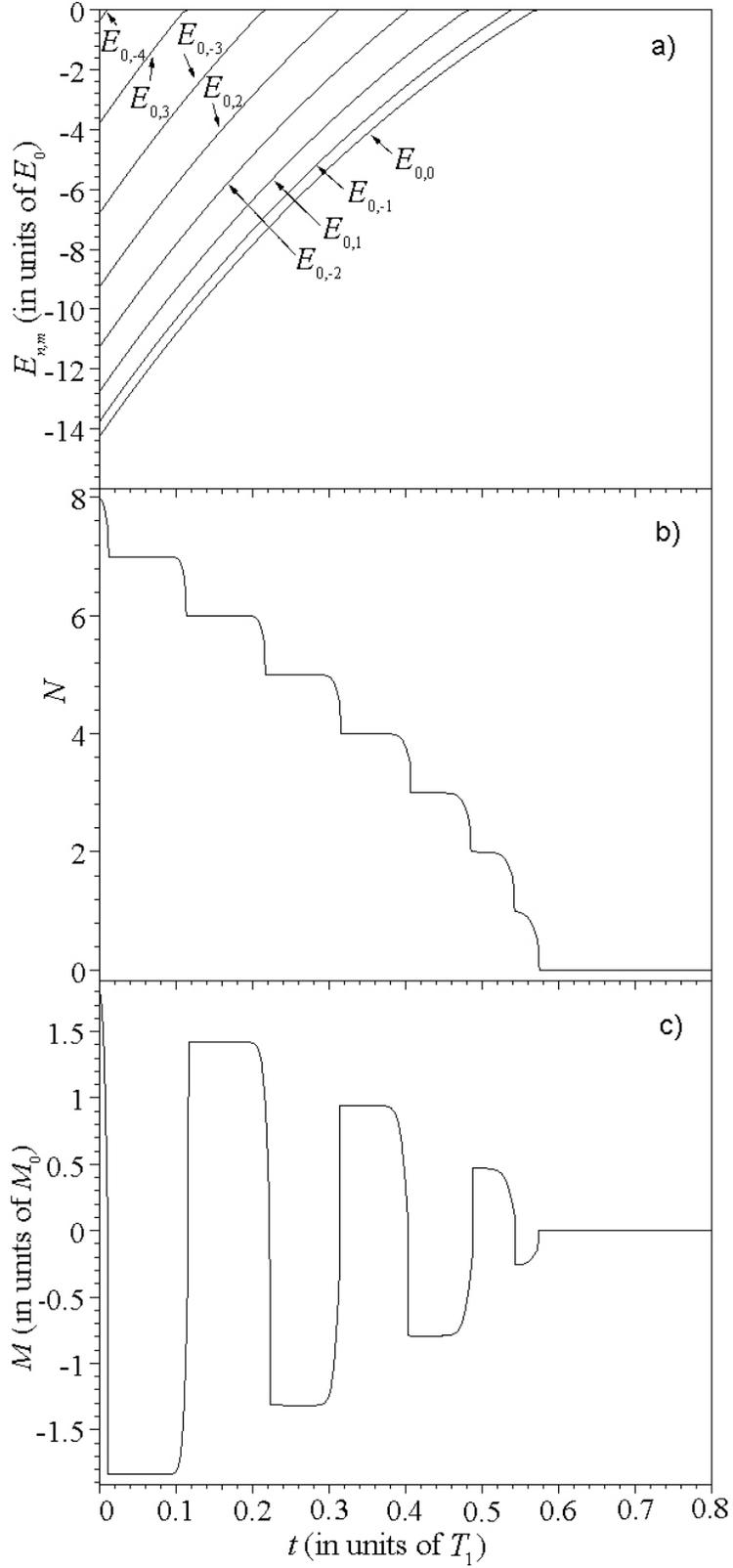}
\bigskip
\caption{Time evolution of energy levels $E_{0,m}$ (a), the number
of levels in NSPI QR $N$ (b), and the magnetic moment $M$ of the
NSPI QR (c) for $T_{1}/t_{0}=0.1$, $kT/E_{0}=0.1$, $\hbar
\protect\omega _{c}/E_{0}=1$, $V_{0}(t=0)/E_{0}=-60$,
$r_0=1\protect\mu m$, $d=0.1\protect\mu m$.}
\end{figure}

\end{document}